\documentclass[aps,pra,reprint,superscriptaddress,nofootinbib]{revtex4-1}

\usepackage{hyperref}
\usepackage{graphicx}
\usepackage{amsmath}
\usepackage{tikz}

\usepackage[xcolor,final]{changes}

\definechangesauthor[color=blue]{VS}
\definechangesauthor[color=magenta]{NL}
\definechangesauthor[color=magenta]{NLnew}
\definechangesauthor[color= orange]{GS}
\definechangesauthor[color=cyan]{ABU}

\newcommand{\ket}[1]{|#1\rangle}
\newcommand{\bra}[1]{\langle#1|}

\begin{document}

\title{Protecting entanglement of twisted photons by adaptive optics}
\author{Nina Leonhard}
\affiliation{Fraunhofer Institute for Applied Optics and Precision Engineering, Albert-Einstein-Stra\ss{}e 7,
07745 Jena, Germany}
\author{Giacomo Sorelli}
\author{Vyacheslav N. Shatokhin}
\affiliation{Physikalisches Institut, Albert-Ludwigs-Universit\"at Freiburg, Hermann-Herder-Stra\ss{}e 3, D-79104 Freiburg, Germany}
\author{Claudia Reinlein}
\affiliation{Fraunhofer Institute for Applied Optics and Precision Engineering, Albert-Einstein-Stra\ss{}e 7,
07745 Jena, Germany}
\author{Andreas Buchleitner}
\affiliation{Physikalisches Institut, Albert-Ludwigs-Universit\"at Freiburg, Hermann-Herder-Stra\ss{}e 3, D-79104 Freiburg, Germany}

\date{\today}

\begin{abstract}

We study the efficiency of adaptive optics (AO) correction for the free-space propagation of entangled photonic orbital-angular-momentum (OAM) qubit states, to reverse moderate atmospheric turbulence distortions. 
We show that AO can significantly reduce crosstalk to modes within and outside the encoding subspace   and thereby stabilize entanglement against turbulence. This method establishes a reliable quantum channel for OAM photons in turbulence, and enhances the threshold turbulence strength for secure quantum communication at least by a factor two.  
\end{abstract}

\maketitle

\setremarkmarkup{(#2)}

\section{Introduction}
\label{sec:intro}
Spatial excitations of the electromagnetic field carrying orbital angular momentum (OAM) \cite{allen92}, often referred to as twisted photons \cite{Molina-Terriza:2007zr}, can be used to encode high-dimensional (entangled) quantum states \cite{Krenn29042014}. 
Not only are these states of fundamental interest \cite{OAMbook}, but also in practice \cite{1367-2630-17-3-033033}, since they can enhance the security of quantum cryptography \cite{PhysRevLett.85.3313,PhysRevA.64.012306} in free space.
However, upon transmission  across atmospheric turbulence,  refractive index fluctuations  are imparted  on  the photons' phase fronts which encode the quantum information \cite{Paterson2005}.
 Whereas successful classical communication with OAM beams has been demonstrated over 143 km \cite{Krenn15112016}, the long-distance transmission of single OAM photons through the atmosphere is more demanding. So far, quantum key distribution over up to 300 m \cite{vallone14,QKDOttawa}, and entanglement distribution over 3~km \cite{Krenn02112015} have been reported. It was suggested  \cite{dunlop17} that to further push the distances  of quantum communication, one has to resort to phase front corrections by  methods of adaptive optics (AO).

Adaptive optics is a well-established scientific discipline and technology that allows to measure and partially correct turbulence-induced errors in astronomy, as well as in  classical free-space optical communication \cite{milonni:476,tysonAO}. A crucial part of any AO system is a circuit connecting the output of the wavefront measurements with a deformable mirror composed of a finite set of electrically controlled segments. By adapting the optical surface of the deformable mirror, it is possible to compensate for phase distortions introduced by turbulence.  Recently, AO has been successfully applied to reduce crosstalk of classical  OAM-multiplexed beams \cite{Ren2014,Ren2014a,Rodenburg2014,Li2014}. 

In this contribution, we evaluate  the potential of AO to mitigate entanglement degradation of photonic OAM states in a moderately turbulent atmosphere \cite{Smith2006,Gopaul2007,Pors:11,PhysRevA.88.012312,Leonhard2015,PhysRevA.92.012326}. The decay of entanglement occurs due to turbulence-induced crosstalk among the OAM modes encoding information. Besides, crosstalk with OAM modes outside the encoding subspace strongly attenuates the detected signal strength.  
As we show below, by compensating the turbulence-induced phase errors, AO counteracts  crosstalk and thereby is able to significantly enhance entanglement, as well as the number of received photons.

The manuscript is organized as follows. In Sec. \ref{sec:model} we present our theoretical model and the details of our numerical simulations. Section \ref{sec:results} contains the results of this work: the protection of entanglement of twisted photons by AO is demonstrated in Sec. \ref{sec:results_a} and the suppression of the qubit error rate -- in Sec. \ref{sec:results_b}. 
Section \ref{sec:conc} concludes our paper.

\begin{figure}
\includegraphics[width=0.9\linewidth]{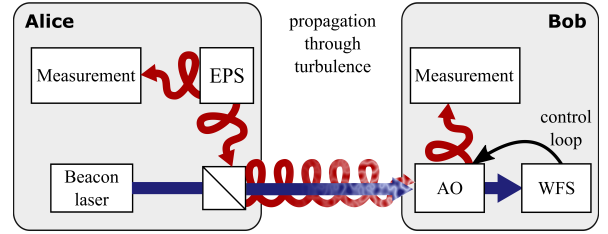}%
\caption{(color online) Sketch of the setup: An entangled photon source (EPS) in Alice's lab produces a pair of entangled twisted photons. One photon is kept in Alice's lab, while the other is sent to Bob, through a free-space channel. A Gaussian beacon (blue arrow) travels along the same path as the twisted photon (red spiral arrow). The AO system mitigates phase distortions of the twisted photons based on the wavefront sensor (WFS) measurements of the beacon. }
\label{fig:Setup}
\end{figure}

\section{Model}
\label{sec:model}
\subsection{Setup}
\label{sec:setup}

Let  us start with the setup here considered, shown in Fig.~\ref{fig:Setup}.
In Alice's laboratory, a biphoton is generated in a maximally entangled (Bell) OAM qubit state, e.g.,
\begin{align}
|\Phi_0\rangle=\frac{1}{\sqrt{2}} \left( |-l_0,l_0 \rangle + |l_0,-l_0 \rangle \right) \, ,
\label{in_Bell}
\end{align}
where $|\pm l_0\rangle$ denotes a single photon state of a Laguerre-Gauss (LG) mode  $LG_{0,\pm l_0}({\bf r},0)$ \cite{allen92} with radial index 0 and azimuthal index $\pm l_0$, at $z=0$ (${\bf r}$ is the transverse coordinate). The constituent photons thus carry an OAM of either $+ \hbar l_0$ or $- \hbar l_0$  \cite{allen92}. 
We assume a typical scenario \cite{Krenn02112015} in which one of the photons stays in Alice's laboratory, while the other one is sent to Bob, via a free-space link of length $L$.
The first photon remains in its initial state, in contrast to the second photon which experiences turbulence-induced distortions. 

\subsection{Evolution of quantum states}
\label{secLquantum}
The evolution under these distortions, for a particular realization of density variations of the medium, can be described by a unitary operator $U_{turb}(L)$ \cite{1751-8121-47-19-195302}, such that propagation of single photon states $\ket{\pm l_0}$ across a turbulent layer  is given by $|\psi_{\pm l_0}\rangle = U_{turb}(L)\ket{\pm l_0}$. The photon at Alice's disposal is not affected by turbulence and we thus act with the identity operator $\openone$ thereupon to obtain the biphoton output state
\begin{subequations}
\begin{align}
&\ket{\Phi}=\{\openone\otimes U_{turb}(L)\}\ket{\Phi_0}, \label{ketturb}\\
&\ket{\tilde{\Phi}} = \{\openone\otimes U_{AO}U_{turb}(L)\}\ket{\Phi_0} , \label{ketao}%
\end{align}%
\label{eq:finalstateturb}%
\end{subequations}
in the absence and in the presence of AO correction, respectively.
In our simulation, we evaluate the unitary operator $U_{turb}(L)$ implicitly, by connecting the mode functions of the input and output states using the extended Huygens-Fresnel principle (see Appendix \ref{appendA}).

Since we have no interest in the specific realizations of turbulence, we need to perform a disorder average of the output biphoton states over different realizations, to obtain the mixed state
\begin{align}
\sigma&=\Big\langle\ket{\Phi}\bra{\Phi}\Big\rangle ,
\label{eq:finalDM} 
\end{align}
where  $\langle\cdots\rangle$ denotes the disorder average. 
Finally, Bob's photon is projected onto the encoding subspace, while Alice's photon already is in this subspace. We describe this procedure by the operator $\openone\otimes \Pi_0$, where
$\Pi_0=|-l_0\rangle \langle -l_0 |+|l_0 \rangle \langle l_0 |$.
The disorder averaged projected biphoton state thus reads
\begin{equation}
\rho=(\openone\otimes\Pi_0)\sigma(\openone\otimes\Pi_0)/\mathcal{N},
\end{equation}
where the factor $\mathcal{N} = {\rm tr} \{(\openone\otimes\Pi_0)\sigma\}$ is required for renormalization. 
We recall that an average trace $\mathcal{N}<1$ of the density matrix
indicates losses which may render quantum communication impossible.

In a final step, we have to evaluate the disorder averaged output state's entanglement. This can be quantified via concurrence \cite{Wootters1998}
\begin{equation}
C = \mathrm{max} \left\lbrace \sqrt{\lambda_1} - \sqrt{\lambda_2} -\sqrt{\lambda_3} - \sqrt{\lambda_4}, 0 \right\rbrace, \label{eq:Wootters}
\end{equation}
where the $\lambda_i$ are the eigenvalues, in decreasing order, of the matrix $\mathcal{R} = \rho (\sigma_y \otimes \sigma_y) \rho^* (\sigma_y \otimes \sigma_y)$, and $\sigma_y$ denotes the second Pauli matrix.
When AO compensation is employed, we simply need to replace $\ket{\Phi}$ with $\ket{\tilde{\Phi}}$ in Eq. \eqref{eq:finalDM}.

In the following, we present the details of the numerical simulation of the atmospheric channel and of the adaptive optics system. 

\subsection{Numerical simulation details}
\subsubsection{Multiple phase screen method}
In our numerical simulations, we implemented the extended Huygens-Fresnel principle via a multiple phase screen approach. Therein, three-dimensional turbulence is described by equally spaced thin phase screens. Each screen introduces random phase distortions in accordance with Kolmogorov turbulence theory \cite{andrews} and the beam experiences free diffraction in vacuum between the screens.    
The  random phase screens were generated using the Kolmogorov spectral density $\Phi_n(\kappa) = 0.033C_n^2\kappa^{-11/3}$, where $\kappa$ is the spatial wave vector in the transverse plane and $C_n^2$ is the turbulence phase structure constant \cite{andrews,schmidt}.
Furthermore, the vacuum propagation between the phase screens was performed with a Fresnel propagator \cite{Bakx2002}, while the phase screens were obtained by the subharmonic method using seven subharmonic orders \cite{Lane1992}.

 For the final simulations, we chose four phase screens -- to properly account for moderate scintillation, while we used a single screen for validation of our numerical procedure (see next section). 
The number of phase screens was determined by requiring for each partial propagation step to have a Rytov variance $\sigma_R^2<0.5$ \footnote{This requirement is stricter than that of  $\sigma_R^2<1$ given in Ref.~\cite{andrews}.} for the chosen propagation distance of $L=500$~m and range of $C_n^2$ values. We simulated 19 values of the turbulence strength  $C_n^2$ between $1.4 \times 10^{-15}$m$^{-2/3}$ (weak turbulence) and $1.5 \times 10^{-13}$m$^{-2/3}$ (moderate turbulence), while each data point was obtained from averaging over 1000 realizations of the turbulent phase screens.
Given $C_n^2$, $L$, and $k=2\pi/\lambda$, one can obtain the transverse turbulence correlation length or Fried parameter $r_0 = (0.423 C_n^2 k^2 L)^{-3/5}$ \cite{andrews} which fixes the  turbulence strength $w_0/r_0$. In addition, we assume a telescope diameter of 0.2~m at Bob's receiver which is a reasonable aperture size available both as a  lens or mirror telescope \added[id=GS]{\cite{Krenn02112015}}. The large diameter ensures that most photons are received despite \replaced[id=GS]{diffraction and turbulence-induced broadening and wandering of the light beam}{many turbulence distortions}.
Based on these simulation parameters, we calculate the biphoton output state in the $\pm l_0$ subspace from the overlap between the received field and the non-perturbed initial OAM modes.
All calculations were carried out on a 0.4~m wide grid with $512\times512$ points for a wavelength of 1064~nm
\footnote{The atmosphere is transparent \cite{andrews} and there exist sources of entangled photon pairs \cite{Magnitskiy2015}, at this wavelength. Furthermore, by a proper rescaling the turbulence strength and the propagation parameters our results can be generalized to other wavelengths.}
 and an initial beam waist of the OAM beams of $w_0=0.03$~m.
Since the extent of the LG intensity profile increases as $\sqrt{l_0+1}$ with $l_0$ ($p=0$) \cite{andrews}, we chose a beam waist for the beacon to be $2.45 \,w_0$. Thereby, we ensured an overlap with the intensity profiles of all simulated OAM modes (up to $l_0=5$).

\subsubsection{Single-phase-screen validation}
\begin{figure}\centering
\includegraphics[width=.72\linewidth]{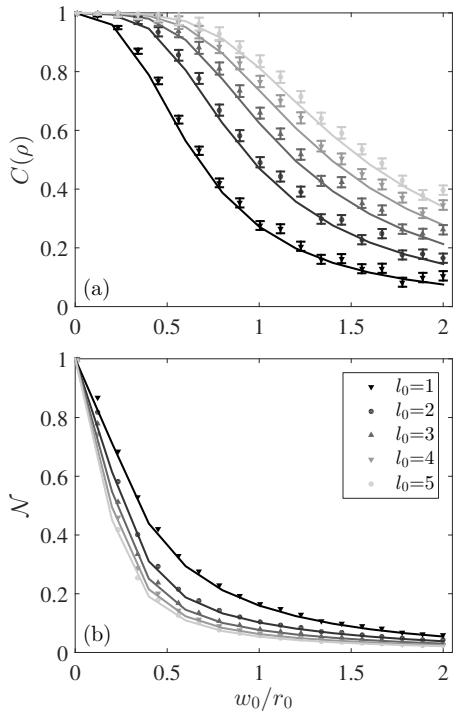}
\caption{Validation of the simulation routine with a single phase screen, (a) concurrence and (b) trace of the final density matrix. Solid lines correspond to analytical results while points refer to numerical data. Error bars in (a) are obtained via error propagation from the standard deviation of the mean on the elements of $\rho$ (see Appendix \ref{appendB}). In (b), the standard deviation of the mean is smaller than the size of the symbols.}
\label{fig:1PS}
\end{figure}

To validate our numerical routine, we first simulated the propagation with a single phase screen, in which case analytical results exist \cite{Leonhard2015} \added[id=GS]{and agree with the earlier theoretical \cite{Smith2006} and experimental \cite{PhysRevA.88.012312} studies of concurrence decay in atmospheric turbulence}. 
Figure~\ref{fig:1PS} shows good agreement both for the concurrence (a) and the trace (b), where solid lines correspond to the analytical theory and points to our numerical data. 

\subsection{Adaptive optics system}
\label{sec:ao}
To measure and correct turbulence-induced distortions, we propose to use an AO system  which consists of a beacon laser, a wavefront sensor and corrective elements such as  deformable or tip/tilt mirrors. In state-of-the-art AO systems \cite{tysonAO}, the time required to perform phase measurements and adjust the mirror into a new position is shorter than the typical timescale of atmospheric changes \cite{milonni:476}, which allows us to neglect the dynamics of the atmosphere.  The classical beacon beam (typically, a Gaussian laser beam \cite{Ren2014}) is sent prior to and along the same path as the quantum light.  
Therefore, we can use its phase, $\varphi_B({\bf r})$, extracted via the wave-front sensor, to correct the phase distortion imprinted onto the single photons. 
Formally, we can express the action of AO by a unitary operator to find the corrected single photon state $|\tilde{\psi}_{\pm l_0}\rangle = U_{AO}U_{turb}(L)\ket{\pm l_0}$ at Bob's receiver. The mode function associated with this corrected state is given by $\tilde{\psi}_{\pm l_0}({\bf r},L) =\exp\{-i\varphi_B({\bf r})\} \psi_{\pm l_0}({\bf r},L)$, where $\psi_{\pm l_0}({\bf r},L)$ is the mode function associated with the state $|\psi_{\pm l_0}\rangle$.

\begin{figure*}
\includegraphics[width=\linewidth]{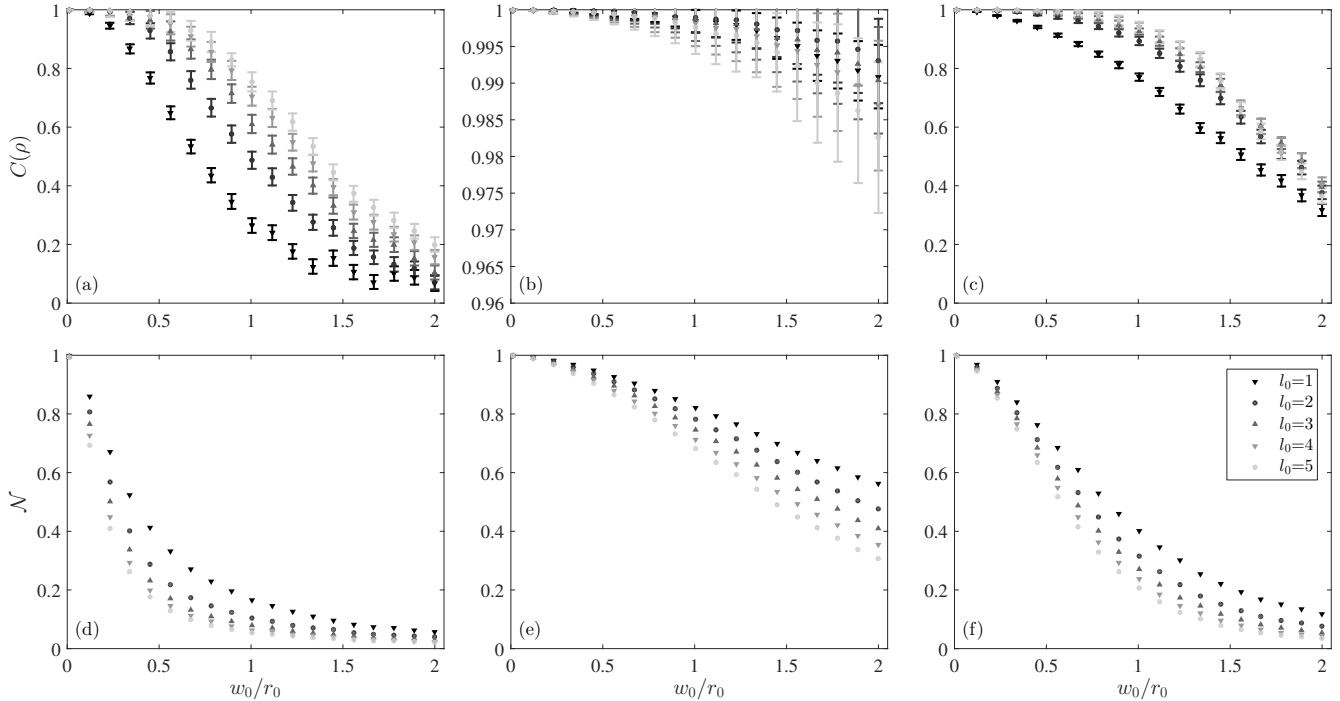}
\caption{Concurrence $C$ \eqref{eq:Wootters}, (a)-(c) and trace $\mathcal{N}$, (d)-(f), of the disorder averaged and projected biphoton output state $\rho$ plotted against the effective turbulence strength $w_0/r_0$, for different degrees of adaptive optics compensation: (a,d) without compensation, (b,e) ideal and (c,f) tip/tilt AO. Note the different scale of the $y$-axis in (b). Error bars in (a)-(c) are obtained via error propagation from the standard deviation of the mean on the elements of $\rho$ (see Appendix \ref{appendB}). In (d)-(f) the standard deviation of the mean is smaller than the size of the symbols.
}
\label{fig:concurrence-trace}
\end{figure*}

The evaluation of the phase $\varphi_B({\bf r})$, and hence of $U_{AO}$ in Eq. \eqref{ketao}, is based upon two ways of  modeling the AO system. 
The first assumes an ideal system able to sense the phase $\varphi_{B,\infty}({\bf r})$ of the beacon field with arbitrary resolution, and to  adapt the deformable mirror's surface correspondingly. The second assumes the simplest  AO  possible which corrects only for a tilt of the wavefront, with respect to the receiver plane.  
In an experiment, this minimalistic scenario is achieved by a single flat mirror which can rotate along both axes perpendicular to the propagation direction -- a so-called tip/tilt (TT) mirror. 
Our calculation of the required mirror rotation, similar to a typical experimental implementation, is based on the Fourier-transforming properties of an ideal lens. Accordingly, a tilted input field is transformed into a displaced focal spot. 
The center of mass of the focal plane intensity thus determines the rotation of the mirror, and thereby, $\varphi_{B,TT}({\bf r})$ \cite{tysonAO}. 
 These optimal and minimalistic AO scenarios allow us to establish an upper and a lower bound for the performance of a real AO system hereafter.

\section{Results}
\label{sec:results}
\subsection{State's entanglement and trace}
\label{sec:results_a}
With the above premises, we can now assess the potential of AO for state and entanglement transmission.

The top row of Fig.~\ref{fig:concurrence-trace} displays
our results for the entanglement evolution under turbulence, without (a) and with optimal (b) or minimalistic (c) AO compensation. 
Figure~\ref{fig:concurrence-trace}(a) establishes the well-known result that, the larger the initial OAM value $l_0$, the more robust entanglement is against (weak and moderate) turbulence \cite{Leonhard2015,PhysRevA.88.012312}. 
Figure~\ref{fig:concurrence-trace}(b) shows that the ideal AO dramatically enhances the output state concurrence, to the extent of being almost fully preserved even in moderate turbulence. 
It might still be surprising that our idealized AO cannot completely recover the initial concurrence. To understand this, we need to consider that diffraction transforms phase distortions into intensity fluctuations, so-called scintillation \cite{andrews}.
It then becomes clear that phase-only AO compensation cannot correct for such intensity distortions and is therefore most efficient for weak to moderate scintillation, i.e. for medium propagation distances and moderate turbulence strengths.

Furthermore, ideal correction inverts the trend observed in Fig.~\ref{fig:concurrence-trace}(a), providing slightly better stability to OAM modes with smaller $l_0$. As for tip/tilt correction, see Fig.~\ref{fig:concurrence-trace}(c), all curves but for $l_0=1$ collapse approximately onto one line for $w_0/r_0\gtrsim 1.5$. Both these observations from (b) and (c) suggest that AO is less effective for higher order OAM modes. 
 We believe that this is due to the different beam geometries  of the OAM modes  and of the Gaussian mode beacon laser, respectively. 
The OAM modes have ring-like intensity patterns with vanishing intensity at the optical vortex -- where the Gaussian beacon has its maximum intensity. Furthermore, OAM modes have a broader intensity profile  which increases with $\sqrt{l_0+1}$ \cite{Paterson2005} while the Gaussian beacon's intensity is essentially localized within a fixed area leading to a decreasing overlap of beacon and OAM beam with increasing $l_0$. 
To reduce this effect, we have chosen a 2.45 times larger beam waist for the beacon than $w_0$ in all of our presented results to ensure an overlap with all modes up to $l_0=5$.
A more quantitative analysis of these geometry-induced effects requires  further optimization of the beacon parameters, and a more detailed
adaptive optics system design which is beyond the scope of our present contribution.

\begin{figure*}%
\includegraphics[width=\linewidth]{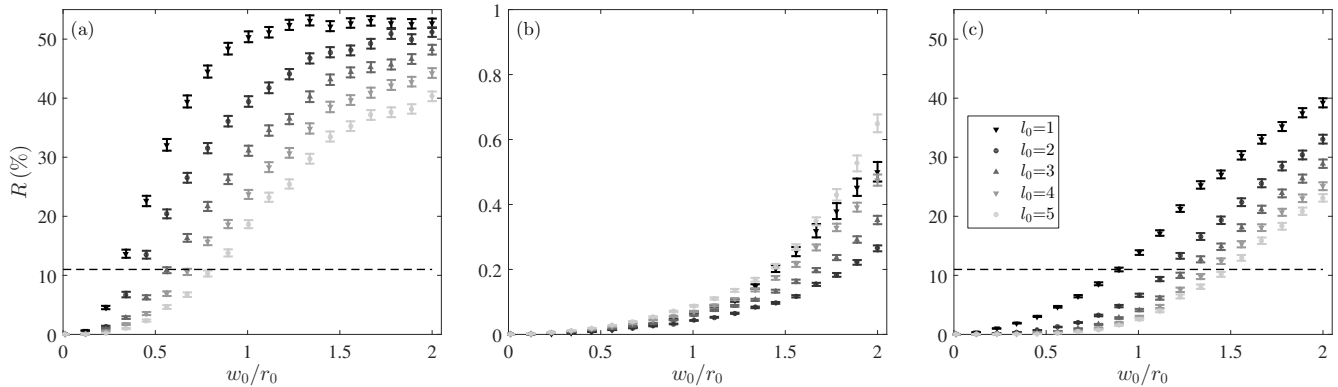}%
\caption{Crosstalk-induced detection error contribution $R$, see Eq.~\eqref{eq:QBER}, to the quantum bit error rate as a function of $w_0/r_0$,  (a) without turbulence compensation, (b) with an ideal, and (c) with tip/tilt AO. Dashed horizontal lines in (a,c) indicate the security threshold of 11\% \cite{QuantCryptReview} (note the different scale of the $y$-axis in (b)). Error bars represent the standard deviation of the mean.}%
\label{fig:QBER}%
\end{figure*}

The bottom row of Fig.~\ref{fig:concurrence-trace} quantifies the loss of the trace ${\mathcal N}$ of the averaged output state's  density matrix as a consequence of the turbulence-induced crosstalk with OAM modes different from $\pm l_0$ \cite{Anguita:08}.  We find that both, ideal (e) and  tip/tilt (f), AO lead to a noteworthy enhancement of the  trace as compared to the uncompensated case (d). Consequently, the number of photons lost due to \replaced[id=GS]{scattering outside the encoding subspace}{crosstalk} can be reduced, which increases the signal-to-noise ratio.  For example, at $w_0/r_0=1$, tip/tilt compensation increases the  trace by a factor between 2 and 4, and ideal AO even achieves  factors between 5 and 13, depending on $l_0$. Interestingly,  higher-order OAM modes exhibit a stronger \added[id=NL]{relative} trace enhancement as compared to  lower-order modes, both for ideal  and tip/tilt AO. Consequently, the number of detectable photons is increased also in  higher-order OAM modes which are more sensitive to crosstalk. AO could thus enable studies of entanglement transmission in state spaces larger than those demonstrated to date.

Let us finally discuss why the efficiency of AO is different for the state's entanglement as compared to its trace. As already mentioned, turbulence causes not only phase, but also intensity fluctuations, which cannot be compensated by AO. Residual intensity fluctuations lead to crosstalk and population of OAM  modes inside and outside the encoding subspace, respectively, even in the case of ideal AO correction. The low residual crosstalk between the modes $\pm l_0$ results in a weak reduction of concurrence. In contrast, small populations in each of the modes outside the encoding subspace result in a relatively large cumulative effect on the trace of the final state. Additionally, the finite receiver aperture could enhance photon losses.

\subsection{Qubit error rate}
\label{sec:results_b}
We finally address an application of our findings in the context of quantum cryptography, where the security of the communication channel is of particular importance. To judge whether an eavesdropper may have gained enough information to render communication insecure, Alice and Bob can evaluate the quantum bit error rate (QBER) on a subset of their measurements \cite{QuantCryptReview}.
In the case of entangled OAM states, intermodal crosstalk is a source of detection errors contributing to the QBER. 
Other, OAM-unrelated, effects, such as detector efficiency and noise statistics, can also contribute to the QBER of the communication channel \cite{QuantCryptReview}. We here restrict our calculations to the detection error rate $R$ caused by crosstalk, 
\begin{equation}
R = \frac{1}{\mathcal{N}} \Big\langle|\bra{+l_0}U_{turb}\ket{-l_0}|^2 + |\bra{-l_0}U_{turb}\ket{+l_0}|^2 \Big\rangle.
\label{eq:QBER}
\end{equation}
It is well-known \cite{QuantCryptReview} that the QBER and, thus, $R$ has to remain below 11\% for secure communication.

Figure \ref{fig:QBER} shows the dependence of $R$ on the turbulence strength, in the absence of AO (a), for ideal AO (b), and for tip/tilt AO (c). Without AO, $R$ quickly prevents secure communication. With tip/tilt correction, the security threshold can be shifted to approximately two times larger turbulence strengths. A dramatic improvement is achieved by ideal AO, with $R<1\%$, such that secure communication can be achieved in the entire range of turbulence strengths considered here, provided all other contributions to QBER remain small enough.

\section{Conclusion}
\label{sec:conc}
In summary, we studied the efficiency of adaptive optics in preventing the loss of entanglement and of norm of OAM qubit states in atmospheric turbulence. Whereas without AO compensation both concurrence and trace rapidly decay with increasing turbulence strength, even minimalistic (tip/tilt) correction allows for an enhancement of the latter quantities by a factor of two to four. 
These results suggest that state of the art AO systems \cite{PetitAO2016},  able to correct higher order aberrations of the wave front, bear the potential to enhance these factors still further -- up to the almost complete restoration of entanglement and an increase of the trace by a factor between five and thirteen, in the ideal case.

While technically more involved in theory as well as in experiment, there is no fundamental obstacle to port the here described method to higher-dimensional OAM-entangled states. Furthermore, we believe that with higher-dimensional states we can push quantum communication protocols based on OAM states to longer propagation distances and worse turbulence condition, since the security threshold increases with increasing the dimensionality of the states \cite{QuantCryptReview}.
Likewise, our results imply that AO methods hold some promise to improve the fidelity of other quantum information protocols which suffer from mode distortions by uncontrolled errors \cite{welte17}.

\begin{acknowledgments}
G.S., V.N.S. and A.B. acknowledge support by Deutsche Forschungsgemeinschaft under grant DFG BU 1337/17-1.
\end{acknowledgments}

\appendix
\section{Derivation of the output density matrix}
\label{appendA}
We employ the {\it extended Huygens-Fresnel principle} \cite{andrews,Ren2014} to describe field propagation across a turbulent medium. According to this principle, 
the mode function of the output, single photon, state $\ket{\psi_{\pm l_0}} = U_{turb}(L)\ket{\pm l_0}$ is expressed by an integral,
\begin{equation}
\psi_{\pm l_0}({\bf r},L) 
=  \int d^2r^\prime h({\bf r},{\bf r}^\prime,L) LG_{0\pm l_0}({\bf r}^\prime,0),
\label{final_state}
\end{equation}
where $h({\bf r},{\bf r}^\prime,L) = \bra{{\bf r}, L}U_\text{turb}(L)\ket{{\bf r}^\prime, 0}$  is the spatial response function which incorporates scattering-in-turbulence and diffraction effects.

On the other hand, the single photon state $|\psi_{\pm l_0}\rangle$ can be  expanded in the OAM basis as,
\begin{equation}
|\psi_{\pm l_0}\rangle=\sum_{pl} g^{\pm l_0}_{pl}|pl\rangle,
\end{equation}
where $|pl\rangle$ denotes a single photon state of an $LG_{pl}$ mode and the coefficients $g^{\pm l_0}_{pl}$ are given by the overlap integral,
\begin{equation}
g^{\pm l_0}_{pl}=\int d^2r\,\psi_{\pm l_0}({\bf r},L)LG^*_{pl}({\bf r},L).
\end{equation}
\added[id=NL]{In our numerical calculations, the integral in the above equation is replaced by a finite sum over the pixels in our calculation grid. }
The biphoton output state in Eq.~(\ref{ketturb}), postselected in the encoding subspace, can be expressed in the OAM basis as
\begin{equation}
(\openone \otimes \Pi_0)\ket{\Phi} =\frac{1}{\sqrt{2}} (a\ket{l_0,l_0}\! +\! b \ket{\! -\! l_0,l_0} \!+\! c  \ket{ l_0,\!-l_0}\! +\! d \ket{\!-l_0,\! - l_0}),
\label{phi}
\end{equation}
where the expansion coefficients read 
\begin{subequations}
\begin{align}
a&:=\langle l_0|U_{turb}(L)|-l_0\rangle =g^{-l_0}_{0l_0}\, , \label{eq:a} \\
b&:=\langle l_0|U_{turb}(L)|l_0\rangle=g^{l_0}_{0 l_0}\, ,\label{eq:b} \\
c&:=\langle-l_0|U_{turb}(L)|-l_0\rangle =g^{-l_0}_{0-l_0}\, ,\label{eq:c} \\
d&:= \langle-l_0|U_{turb}(L)|l_0\rangle=g^{l_0}_{0-l_0}\,. \label{eq:d} 
\end{align}
\label{system}
\end{subequations}
By definition, $b$ and $c$ are the survival amplitudes, whereas $a$ and $d$ -- the crosstalk amplitudes \cite{Leonhard2015}. 
In terms of these quantities, the average density matrix in Eq.~(\ref{eq:finalDM}) is given by
\begin{align}
\rho&=\frac{1}{\mathcal{N}}\begin{pmatrix}\langle |a|^2\rangle & \langle a^*b\rangle & \langle a^*c\rangle & \langle a^*d \rangle \\ \langle a^*b\rangle & \langle |b|^2\rangle & \langle b^*c\rangle & \langle b^*d \rangle \\ \langle c^*a \rangle & \langle  c^* b \rangle & \langle  |c|^2\rangle & \langle c^*d \rangle \\ \langle d^*a\rangle & \langle d^*b\rangle & \langle d^*c\rangle & \langle |d|^2 \rangle \end{pmatrix} , \label{eq:finalDMabcd}
\end{align}
with the normalization constant $\mathcal{N}=\langle|a|^2+|b|^2+|c|^2+|d|^2\rangle$.
Using Eqs.~(\ref{system}a-d), we can also write the QBER in Eq.~(\ref{eq:QBER}) as
\begin{equation}
R =\langle|a|^2 + |d|^2\rangle/\mathcal{N}.
\label{R}
\end{equation}
All previous expressions concerned wave propagation without the AO compensation, but they can easily be adapted to the case when AO correction of the phase front is present. Indeed, the mode function for the AO-compensated  states $\ket{\tilde{\psi}_{\pm l_0}}$ is given by
\begin{equation}
\tilde{\psi}_{\pm l_0}({\bf r},L)  = e^{-i \varphi_B({\bf r})} \psi_{\pm l_0} ({\bf r},L), 
\end{equation}
where $\varphi_B({\bf r})$ is either $\varphi_{B,\infty}({\bf r})$ (ideal phase correction) or $\varphi_{B,TT}({\bf r})$ (tip/tilt correction).
Furthermore, the AO-compensated state reads
\begin{equation}
|\tilde{\psi}_{\pm l_0}\rangle=\sum_{pl} \tilde{g}^{\pm l_0}_{pl}|pl\rangle,
\end{equation}
with
\begin{equation}
\tilde{g}^{\pm l_0}_{pl}=\int d^2r \,\psi_{\pm l_0}({\bf r},L)LG^*_{pl}({\bf r},L)e^{-i\varphi_B({\bf r})}.
\label{g_tilde}
\end{equation} 
Analogously, when we expand the biphoton state $|\tilde{\Phi}\rangle$ [see Eq. ~(\ref{ketao})] in the OAM basis, we arrive at equations similar to Eqs.~(\ref{phi})-(\ref{R}).

\section{Error on concurrence}
\label{appendB}
Here we discuss the method to obtain the errors of concurrence through the propagation of the statistical errors of the output density matrix.

Our derivation below follows closely that in Ref.~\cite{Ibrahim2013}, except that we used perturbation theory for non-Hermitian matrices (see below).
First of all, we used the Bloch representation of the density matrix, which renders matrix elements (and consequently, errors) thereof real. In the Bloch representation, the density matrix reads
\begin{equation}
\rho = \sum_{i,j = 0}^3 B_{ij}\sigma_i \otimes \sigma_j,
\label{rho}
\end{equation}
where $B_{ij} = \text{tr}(\rho \, \sigma_i\otimes\sigma_j)$ are the Bloch coefficients, $\sigma_0$ is the $2\times 2$ identity matrix and $\sigma_{1,2,3}$ are the three Pauli matrices. 
Wootters' concurrence is a function of the eigenvalues of the non-Hermitian matrix $\mathcal{R}$ given by
\begin{equation}
\mathcal{R} = \bar{\rho} (\sigma_y \otimes \sigma_y) \bar{\rho}^* (\sigma_y \otimes \sigma_y) ,
\end{equation}
 which can be expressed in the Bloch representation as
\begin{align}
\mathcal{R} &= \sum_{ijkl=0}^3 B_{ij}B_{kl} (\sigma_i \otimes \sigma_j) (\sigma_y \otimes \sigma_y)(\sigma_k \otimes \sigma_l)^*(\sigma_y \otimes \sigma_y) ,\nonumber \\
&=  \sum_{ijkl=0}^3 B_{ij}B_{kl} \Gamma_{ijkl}.
\end{align}
where the four-index tensor $\Gamma_{ijkl} = (\sigma_i \sigma_y \sigma_k^* \sigma_y) \otimes(\sigma_j \sigma_y \sigma_l^* \sigma_y)$. We calculated the error on concurrence by propagating the error on the Bloch coefficients $\Delta B_{ij}$, which was calculated as standard deviation of the mean, assuming $B_{ij}$ to be statistically independent.

Using standard error propagation on Eq.~(\ref{eq:Wootters}), we expressed the error on concurrence as
\begin{equation}
\left( \Delta C \right)^2 = \sum_{i = 1}^4 \left( \frac{\partial C}{\partial \lambda_i} \right)^2 \left( \Delta \lambda_i\right)^2 = \sum_{i=1}^4 \left( \frac{1}{2} \frac{\Delta \lambda_i}{\sqrt{\lambda_i}}\right)^2,
\end{equation}
where $\Delta \lambda_i$ are the errors on the eigenvalues of the matrix $\mathcal{R}$. 
To calculate the errors $\Delta \lambda_i$, we first found the error on the matrix $\mathcal{R}$ as 
\begin{align}
\Delta \mathcal{R}  &= \sum_{mn = 0} ^3 \frac{\partial R}{\partial B_{mn}}\Delta  B_{mn} \nonumber , \\&= \sum_{klmn = 0} ^3 ( B_{kl} \Gamma_{mnkl} + B_{kl}\Gamma_{klmn}) \Delta  B_{mn}.
\label{Delta_R}
\end{align}
Finally, we used perturbation theory for non-Hermitian matrices \cite{JamesPRA2001} to calculate the errors on the eigenvalues as
\begin{equation}
\Delta \lambda_i = W^\dagger_i \Delta \mathcal{R} V_i,
\end{equation} 
where $W_i$ ($V_i$) are the left (right) eigenvectors of $\mathcal{R}$.

\bibliographystyle{apsrev4-1}
\end{document}